\documentclass[12pt,a4paper]{article}
\usepackage{amsmath}
\usepackage{graphicx,amsmath,amssymb}

\usepackage{cite}
\usepackage{graphicx,amsmath,amssymb}

\DeclareMathAlphabet{\mathpzc}{OT1}{pzc}{m}{it}

\numberwithin{equation}{section}

\title{\bf Order-by-order Analytic Solution to the BFKL Equation} 
\author{D.~A.~Ross$^{1}$ and A.~Sabio~Vera$^{2,3}$\\ 
\\
{\small $^1$ School of Physics and Astronomy, Southampton U.,}\\ 
{\small Southampton SO17 1BJ, UK}\\
{\small $^2$ Instituto de F{\' \i}sica Te{\' o}rica UAM/CSIC, Nicol{\'a}s Cabrera 15}\\ 
{\small \& U. Aut{\' o}noma de Madrid, E-28049 Madrid, Spain}\\
{\small $^3$ Physics Department, Theory Unit, CERN,}\\
{\small CH-1211  Gen{\` e}ve 23, Switzerland} }

\begin{document} 

\maketitle 

We propose a regularization of the BFKL equation which allows for its solution in each order of perturbation theory by means of 
a sum over multiple poles. This sum can be presented in a rather simple formula for the Fourier transform in the azimuthal angle of the 
gluon Green function. In order to test our method, we have compared a few orders in the expansion to previous results by Del Duca, Dixon, Duhr and Pennington, finding agreement. Our formalism is general and can be applied to other, more complicated, kernels. 

\section{Introduction}

We start with the following representation in momentum space of the forward BFKL equation at leading order~\cite{BFKL}:
\begin{eqnarray}
\frac{\partial \varphi \left({\bf q}_1,{\bf q}_2,Y\right)}{\overline{\alpha} \partial Y} 
=  \int \frac{d^2 {\bf k}}{\pi} \Bigg\{\frac{\varphi  \left({\bf k},{\bf q}_2,Y\right)}{\left({\bf k}-{\bf q}_1\right)^2}- \frac{{\rm min} (k^2,q_1^2) \varphi \left({\bf q}_1,{\bf q}_2,Y\right)}{\left|k^2-q_1^2\right| \left(k^2+q_1^2\right)}
\Bigg\},  \label{eq1}
\end{eqnarray}
where $$ \overline{\alpha} \ \equiv \ \frac{C_A \alpha_s}{\pi} $$
and the function  $\varphi \left({\bf q}_1,{\bf q}_2,Y\right)$
is the amplitude for the forward scattering of two gluons with virtualities
${\bf q}_1$ and ${\bf q}_2$ respectively and rapidity gap $Y$.
The first term on the RHS of eq.(\ref{eq1}) represents the emission of an extra
gluon  in the $s$-channel with an infinitesimal increase in rapidity, and the 
second term encodes the fact that the gluons emitted in the $t$-channel
are ``reggeized'' so that they have a rapidity gap dependence. These two terms
are individually infrared divergent as can be seen from the integrand
at the point ${\bf k}={\bf q}_1$, but the divergences cancel when these
two components are added together.

The solution to the differential equation (\ref{eq1})
may be written in terms of the perturbative expansion
\begin{eqnarray}
\varphi \left({\bf q}_1,{\bf q}_2,Y\right) &=&  
\delta^{(2)} \left({\bf q}_1-{\bf q}_2\right) + 
\sum_{M=1}^\infty \frac{\left(\overline{\alpha} Y\right)^M}{M!} {\cal F}_M \left({\bf q}_1,{\bf q}_2\right).
\end{eqnarray}
The following relation then holds
\begin{eqnarray}
 {\cal F}_{M+1} \left({\bf q}_1,{\bf q}_2\right) = 
\int \frac{d^2 {\bf k}}{\pi} \Bigg\{\frac{{\cal F}_M \left({\bf k},{\bf q}_2\right)}{\left({\bf k}-{\bf q}_1\right)^2}  - \frac{{\rm min} (k^2,q_1^2) {\cal F}_M \left({\bf q}_1,{\bf q}_2\right)}{\left|k^2-q_1^2\right| \left(k^2+q_1^2\right)} 
\Bigg\},
\end{eqnarray}
where
\begin{eqnarray}
{\cal F}_0 \left({\bf q}_1,{\bf q}_2\right) \equiv \delta^{(2)} \left({\bf q}_1-{\bf q}_2\right) 
= \sum_{n=-\infty}^\infty 
\int \frac{d \gamma}{2 \pi i} \left(\frac{q_1^2}{q_2^2}\right)^{\gamma-1} 
\frac{e^{i n \left(\theta_{q_1}-\theta_{q_2}\right)}}{\pi q_2^2}. 
\label{h0}
\end{eqnarray}
Using the representation for the propagator
\begin{eqnarray}
\frac{1}{\left({\bf q}_1 - {\bf q}_2\right)^2}  =   
 \sum_{n=-\infty}^\infty e^{i n \left(\theta_{q_1}-\theta_{q_2}\right)}  
 \left(\frac{\theta \left(q_1 -q_2\right) \left(\frac{q_2}{q_1}\right)^{|n|} }{q_1^2-q_2^2}+ \frac{\theta \left(q_2 -q_1\right) \left(\frac{q_1}{q_2}\right)^{|n|}}{q_2^2-q_1^2} \right) 
\end{eqnarray}
we can write
\begin{eqnarray}
{\cal F}_{M+1} \left(q_1^2,q_2^2,\theta_{q_1},\theta_{q_2}\right) &=& 
 \int_0^{2 \pi} \frac{d \theta_k}{2 \pi} \sum_{n=-\infty}^\infty e^{i n \left(\theta_{k}-\theta_{q_1}\right)}\\
&&\hspace{-4.6cm} \times 
 \Bigg\{\int_{q_1^2}^\infty  \frac{d k^2 }{k^2-q_1^2} \left[
\left(\frac{q_1}{k}\right)^{|n|} {\cal F}_M \left(k^2,q_2^2,\theta_{k},\theta_{q_2}\right) - \frac{q_1^2}{k^2+q_1^2} {\cal F}_M \left(q_1^2,q_2^2,\theta_{q_1},\theta_{q_2}\right)\right] \nonumber\\
&&\hspace{-4.6cm} + \int_0^{q_1^2}   \frac{d k^2 }{q_1^2-k^2}\left[
\left(\frac{k}{q_1}\right)^{|n|} {\cal F}_M \left(k^2,q_2^2,\theta_{k},\theta_{q_2}\right) - \frac{k^2}{k^2+q_1^2} {\cal F}_M \left(q_1^2,q_2^2,\theta_{q_1},\theta_{q_2}\right)\right]
 \Bigg\}.\nonumber
\end{eqnarray}
Now we introduce a  variable $x$ defined as
\begin{eqnarray} 
x \ \equiv \ {\rm min}\left(\frac{q_1^2}{k^2},\frac{k^2}{q_1^2}\right). 
\end{eqnarray}
The range of integration is $0 \, < \, x \, \leq \, 1$ and the 
infrared divergences occur at $x=1$. We may therefore regularize these
divergences by replacing the denominator term $(1-x)^{-1}$ by $(1-x)^{\epsilon-1}$
and ultimately taking the limit $\epsilon \, \to \, 0 $.

After introducing our regulator $\epsilon$, the previous expression  becomes
\begin{eqnarray}
{\cal F}_{M+1} \left(q_1^2,q_2^2,\theta_{q_1},\theta_{q_2}\right) &=& 
 \int_0^{2 \pi} \frac{d \theta_k}{2 \pi} \sum_{n=-\infty}^\infty e^{i n \left(\theta_{k}-\theta_{q_1}\right)} \int_0^1 dx \left(1-x\right)^{\epsilon-1} \\
&&\hspace{-4.8cm} \times   \Bigg\{x^{\frac{|n|}{2}}
 \left[\frac{1}{x} 
 {\cal F}_M \left(\frac{q_1^2}{x},q_2^2,\theta_{k},\theta_{q_2}\right)
 + {\cal F}_M \left(x q_1^2,q_2^2,\theta_{k},\theta_{q_2}\right)\right]
 -2 {\cal F}_M \left(q_1^2,q_2^2,\theta_{q_1},\theta_{q_2}\right)\Bigg\}. \nonumber
\end{eqnarray}
Making use of the last representation for ${\cal F}_0$ in eq.~(\ref{h0}), we can 
iterate this expression and write
\begin{eqnarray}
{\cal F}_{M} \left(q_1^2,q_2^2,\theta_{q_1},\theta_{q_2}\right) &=& 
\lim_{\epsilon \to 0} 
\sum_{n=-\infty}^\infty 
\int \frac{d \gamma}{2 \pi i} \left(\frac{q_1^2}{q_2^2}\right)^{\gamma-1} 
\frac{e^{ i n \left(\theta_{q_1} - \theta_{q_2} \right)}}{\pi q_2^2}\nonumber\\
&&\hspace{-2cm}\times
 \left(\int_0^1 dx \left(1-x\right)^{\epsilon-1} \left[
 x^{\frac{|n|}{2}} \left(x^{\gamma-1}+ x^{-\gamma} \right)-2\right]\right)^M,
 \label{FMfirstintegral}
\end{eqnarray}
which, after performing the integration in $x$, reads
\begin{eqnarray}    
 {\cal F}_{M} \left(q_1^2,q_2^2,\theta_{q_1},\theta_{q_2}\right) &=& \lim_{\epsilon \to 0} 
\sum_{n=-\infty}^\infty 
\int \frac{d \gamma}{2 \pi i} \left(\frac{q_1^2}{q_2^2}\right)^{\gamma-1} 
\frac{e^{ i n \left(\theta_{q_1} - \theta_{q_2} \right)}}{\pi q_2^2}\nonumber\\
&&\hspace{-3.cm}\times 
\left(\frac{\Gamma\left(\gamma+\frac{|n|}{2}\right)  \Gamma \left(\epsilon\right)}{\Gamma\left(\gamma+\frac{|n|}{2}+\epsilon\right)}
+\frac{\Gamma\left(1-\gamma+\frac{|n|}{2}\right)  \Gamma \left(\epsilon\right)}{\Gamma\left(1-\gamma+\frac{|n|}{2}+\epsilon\right)} -\frac{2}{\epsilon}\right)^M.
\end{eqnarray}
Without loss of generality, we consider the $q_1^2 > q_2^2$ case and 
slightly modify our regularization prescription accordingly (this will simplify the intermediate steps of our calculations without affecting the final result):
\begin{eqnarray}  
{\cal F}_{M} \left(q_1^2,q_2^2,\theta_{q_1},\theta_{q_2}\right) &=&  \lim_{\epsilon \to 0}  
\sum_{n=-\infty}^\infty 
\int \frac{d \gamma}{2 \pi i} \left(\frac{q_1^2}{q_2^2}\right)^{\gamma-1} 
\frac{e^{ i n \left(\theta_{q_1} - \theta_{q_2} \right)}}{\pi q_2^2}\nonumber\\
&&\hspace{-3cm}\times 
\left(\frac{\Gamma\left(\gamma+\frac{|n|}{2}\right)  \Gamma \left(\epsilon\right)}{\Gamma\left(\gamma+\frac{|n|}{2}+\epsilon\right)}
- \psi \left(1-\gamma + \frac{|n|}{2}\right) + \psi (1) -\frac{1}{\epsilon}\right)^M.
\end{eqnarray}   
It is convenient to shift the $\gamma$ variable and use the notation 
$\theta = \theta_{q_1} - \theta_{q_2} $, $\rho = q_2^2 / q_1^2$ to write
\begin{eqnarray}   
{\cal F}_{M}  =  \lim_{\epsilon \to 0} \sum_{n=-\infty}^\infty \frac{e^{ i n \theta}}{\pi q_1^2}
\int \frac{d \gamma}{2 \pi i} \rho^{\frac{|n|}{2}-\gamma} 
 \left(\frac{\Gamma
   \left(\gamma \right)\Gamma (\epsilon ) }{\Gamma \left(\gamma +\epsilon\right)}-H_{|n|-\gamma}   - \frac{1}{\epsilon }\right)^M,
\end{eqnarray}
where $H_{|n|-\gamma} = \psi (1-\gamma +|n|) - \psi (1)$ is the harmonic number.

This expression can be evaluated by moving the contour of integration, which is a vertical 
line in the region $0 < \Re( \gamma) < 1$, to the left of all the poles generated by the powers 
of the gamma function $\Gamma(\gamma)$, as it is usually done in Mellin-Barnes 
integrations. The purpose of our work is to show that this procedure generates the 
correct order-by-order expansion of the gluon Green function. 

In order to get a more explicit expression we first introduce the expansion
\begin{eqnarray}
{\cal F}_{M}  &=&  \lim_{\epsilon \to 0} 
\sum_{n=-\infty}^\infty \frac{e^{ i n \theta} \rho^{\frac{|n|}{2}} }{\pi q_1^2}
 \sum_{L=1}^M \frac{(-1)^{M-L} M!}{L! (M-L)!}  \nonumber\\
&\times& \int \frac{d \gamma}{2 \pi i} 
 \left(\frac{\Gamma
   \left(\gamma \right)\Gamma (\epsilon ) }{\Gamma \left(\gamma +\epsilon\right)}\right)^L 
   \frac{\left(H_{|n|-\gamma} + \frac{1}{\epsilon }\right)^{M-L}}{\rho^\gamma},
\end{eqnarray}
where we have removed the $L=0$ term since it does not generate any pole in the region  
$\Re (\gamma) \leq 0$. 
The sum over residues at $\gamma = - S = 0, -1,-2 \dots$ is
\begin{eqnarray}
{\cal F}_{M}  &=&   \lim_{\epsilon \to 0} 
\sum_{n=-\infty}^\infty 
\frac{e^{ i n \theta}\rho^{\frac{|n|}{2}}}{\pi q_1^2}
\sum_{L=1}^M \frac{\left(- 1\right)^{M-L} M!}{L! (L-1)! (M-L)!} 
\\
&\times&
\sum_{S=0}^\infty 
\frac{d^{L-1}}{d \gamma^{L-1}} \bigg[
\left(\frac{(\gamma+S) \Gamma\left(\gamma\right) \Gamma\left(\epsilon\right)}{\Gamma\left(\gamma+\epsilon\right)}
\right)^{L}
\frac{\left(H_{|n|-\gamma} + \frac{1}{\epsilon }\right)^{M-L}}{\rho^\gamma} \bigg]_{\gamma = -S}.\nonumber
\end{eqnarray}
It can be expressed in the form
\begin{eqnarray}
{\cal F}_{M}  &=&  \frac{1}{\pi q_1^2} 
\sum_{L=1}^M 
\sum_{K=0}^{L-1} 
\sum_{A=0}^{L-1-K}
\frac{\left(- 1\right)^{M-L} M! \left(\ln{\frac{1}{\rho}}\right)^{L-1-K-A} }{L! K! (M-L)! A! (L-1-K-A)!} 
\nonumber\\
&\times& \sum_{n=-\infty}^\infty  \sum_{S=0}^\infty e^{ i n \theta}\rho^{S+\frac{|n|}{2}}
{\cal R}_M^{\left[L,K,A\right]} (n,S,\epsilon)
\label{General}
\end{eqnarray}
with
\begin{eqnarray}
{\cal R}_M^{\left[L,K,A\right]} (n,S,\epsilon) &=& \lim_{\epsilon \to 0} \bigg[\frac{d^{K}}{d \gamma^{K}} 
\left(\frac{(\gamma+S) \Gamma\left(\gamma\right) \Gamma\left(\epsilon\right)}{\Gamma\left(\gamma+\epsilon\right)}
\right)^{L}\bigg]_{\gamma = -S} \nonumber\\
&\times&
\left[\frac{d^{A}}{d \gamma^{A}} \left(H_{|n|-\gamma} + \frac{1}{\epsilon }\right)^{M-L}\right]_{\gamma = -S}.
\end{eqnarray}
For calculational purposes, it is convenient to define
$$ \eta \ = \ \gamma +S $$
and express the coefficients ${\cal R}$ in terms of derivatives w.r.t. $\eta$ at vanishing $\eta$, i.e.
\begin{eqnarray}
{\cal R}_M^{\left[L,K,A\right]} (n,S,\epsilon) &=&
 \nonumber \\ & & \hspace*{-3cm}
 \lim_{\epsilon \to 0} \bigg[\frac{d^{K}}{d \eta^{K}} 
\left((\eta+\epsilon) \Gamma\left(\epsilon\right) 
\frac{\Gamma\left(1-\eta\right)   \Gamma\left(1+\eta\right) }  {\Gamma\left(1-\eta-\epsilon\right)   \Gamma\left(1+\eta+\epsilon\right) }  
\frac{\Gamma\left(S+1-\eta-\epsilon\right)}{\Gamma\left(S+1-\eta\right)}
\right)^{L}\bigg]_{\eta = 0} \nonumber\\
&\times&
\left[\frac{d^{A}}{d \eta^{A}} \left(H_{|n|+s-\eta} + \frac{1}{\epsilon }\right)^{M-L}\right]_{\eta = 0}.
\label{coeffs} \end{eqnarray}

This is the main result of our work. To show how this general formula works~\footnote{In the following we will simply write ${\cal R}_M^{\left[L,K,A\right]}$ instead of ${\cal R}_M^{\left[L,K,A\right]} (n,S,\epsilon)$.} in practice we solve the first three orders in perturbation theory, {\it i.e.}, we calculate ${\cal F}_{M=1,2,3}$. We have checked that the obtained results are in agreement with a numerical implementation of ${\cal F}_M$ as in eq.~(\ref{FMfirstintegral}) and 
with the results presented in~\cite{Dixon}.

\section{First iteration}

Let us evaluate the first term :
\begin{eqnarray}
{\cal F}_{1}  &=&  \sum_{n=-\infty}^\infty \frac{e^{ i n \theta}}{\pi q_1^2}
\int \frac{d \gamma}{2 \pi i} \rho^{-\gamma+\frac{|n|}{2}} 
 \left(\frac{\Gamma
   \left(\gamma \right)\Gamma (\epsilon ) }{\Gamma \left(\gamma +\epsilon\right)}-H_{|n|-\gamma}   - \frac{1}{\epsilon }\right).
\end{eqnarray}
The sum over the residues at $\gamma = - S = 0, -1, \dots$, applying eq.~(\ref{General}), reads
\begin{eqnarray}
{\cal F}_{1}  &=&   \sum_{n=-\infty}^\infty  \sum_{S=0}^\infty \frac{e^{ i n \theta} \rho^{S+\frac{|n|}{2}} }{\pi q_1^2}
{\cal R}_1^{\left[1,0,0\right]}.  
\end{eqnarray}
From eq.~(\ref{coeffs}), we can easily see that
\begin{eqnarray} {\cal R}_1^{\left[1,0,0\right]}  &=&  1 \end{eqnarray}
so that

\begin{eqnarray}
 {\cal F}_{1}
&=&  \frac{1}{\pi q_1^2 \left(1+\rho-2 \sqrt{\rho} \cos{(\theta)}\right)} ~=~  \frac{1}{\pi \left({\bf q}_1 - {\bf q}_2\right)^2},
\end{eqnarray}
in agreement with the result of~\cite{Dixon} \footnote{The definition of ${\cal F}_i$ used here differs from that of 
ref.~\cite{Dixon} by a factor of  $\pi \left({\bf q}_1 - {\bf q}_2\right)^2$.}.

It is interesting to write the result for the azimuthal angle averaged kernel (keeping in mind that $\rho < 1$), which corresponds to $n=0$:
\begin{eqnarray}
\tilde{\cal F}_{1}  &=&  \frac{1}{\pi q_1^2 \left(1-\rho\right)}.
\end{eqnarray}

\section{Second iteration}

Now, we study the second order case:
\begin{eqnarray}
{\cal F}_{2}  &=&  \sum_{n=-\infty}^\infty \frac{e^{ i n \theta}}{\pi q_1^2}
\int \frac{d \gamma}{2 \pi i} \rho^{-\gamma+\frac{|n|}{2}} 
 \left(\frac{\Gamma
   \left(\gamma \right)\Gamma (\epsilon ) }{\Gamma \left(\gamma +\epsilon\right)}-H_{|n|-\gamma}   - \frac{1}{\epsilon }\right)^2.
\end{eqnarray}

The only contributions with a non-zero residue at $\gamma = - S = 0, -1, -2 \dots$ 
can be written as
\begin{eqnarray}
{\cal F}_{2}  =  2 \sum_{n=-\infty}^\infty \frac{e^{ i n \theta}}{\pi q_1^2}
\int \frac{d \gamma}{2 \pi i} \rho^{-\gamma+\frac{|n|}{2}} 
 \frac{\Gamma
   \left(\gamma \right)\Gamma (\epsilon ) }{\Gamma \left(\gamma +\epsilon\right)}
   \left(\frac{\Gamma
   \left(\gamma \right)\Gamma (\epsilon ) }{2 \Gamma \left(\gamma +\epsilon\right)}-H_{|n|-\gamma}   - \frac{1}{\epsilon }\right).
\end{eqnarray}
Making use of the formula in eq.~(\ref{General}) we have
\begin{eqnarray}
{\cal F}_{2}  =  \sum_{n=-\infty}^\infty  \sum_{S=0}^\infty \frac{e^{ i n \theta} \rho^{S+\frac{|n|}{2}} }{\pi q_1^2}\Bigg\{-2 {\cal R}_2^{\left[1,0,0\right]} + {\cal R}_2^{\left[2,1,0\right]} 
+ \left(\ln{\frac{1}{\rho}} \right)  {\cal R}_2^{\left[2,0,0\right]} \Bigg\}.
 \label{ff22}
\end{eqnarray}
From eq.~(\ref{coeffs}) we have
\begin{eqnarray}
{\cal R}_2^{\left[1,0,0\right]} &=&    
 H_{S+|n|}-H_S+\frac{1}{\epsilon };
\end{eqnarray}
\begin{eqnarray}
 {\cal R}_2^{\left[2,1,0\right]}
&=& \frac{2}{\epsilon }-4 H_S;
\end{eqnarray}
and
\begin{eqnarray}
{\cal R}_2^{\left[2,0,0\right]} &=& 1.
\end{eqnarray}
Note that, as expected, the pole terms in $\epsilon$ cancel in
eq.~(\ref{ff22}).
We can write ${\cal F}_2$ in the more explicit form
\begin{eqnarray}
{\cal F}_{2}  &=& 2  \sum_{n=-\infty}^\infty\sum_{S=0}^\infty
\frac{ e^{i n \theta }    }{\pi  {q}_1^2 } \rho^{\frac{|n|}{2}+S} \lim_{\epsilon \to 0}
\frac{(-1)^{S} \Gamma (\epsilon )}{ S! \, \Gamma (\epsilon -S)}
   \nonumber\\
   &\times& \Bigg\{  \frac{(-1)^{S} \Gamma (\epsilon ) }{S! \, \Gamma
   (\epsilon -S)} \left( H_S-H_{\epsilon -1-S}- \frac{\ln
   \rho}{2} \right)- H_{|n|+S}-\frac{1}{\epsilon }
   \Bigg\}.
\end{eqnarray}
Taking the $\epsilon \to 0$ limit before considering the sum  we have
\begin{eqnarray}
{\cal F}_{2}  &=& -  \sum_{n=-\infty}^\infty \frac{e^{i n \theta } \rho ^{\frac{|n|}{2}} }{\pi  q_1^2}\sum_{S=0}^\infty \rho^S 
\left(2   \left(H_S + H_{S+|n|}\right)+\ln{\rho}\right).
\end{eqnarray}
We can now use eq.~(\ref{rhoH}) and eq.~(\ref{rhoHsplusn}) 
to perform the sum over residues:
\begin{eqnarray}
{\cal F}_{2}  =   \sum_{n=-\infty}^\infty 2 \frac{e^{i n \theta }
    \rho ^{\frac{|n|}{2}}}{\pi  q_1^2 (1-\rho)} \Bigg\{
     \ln (1-\rho )  - \frac{\ln (\rho )}{2}
     - H_{|n|} - \sum_{m=1}^\infty \frac{\rho^m}{m+|n|}  
    \Bigg\}.
\end{eqnarray}
The sum over $n$ has several terms which can be calculated using eq.~(\ref{F2n1}), eq.~(\ref{F2n2}) 
and  eq.~(\ref{F2n3}). The final result is 
\begin{eqnarray}
{\cal F}_{2} 
&=& \frac{2 \left(\ln \left(1+\rho -2 \sqrt{\rho } \cos (\theta
   )\right)-\frac{\ln (\rho )}{2}\right)}{\pi 
   q_1^2 \left(1+\rho-2 \sqrt{\rho } \cos (\theta )\right)} \nonumber\\
   &=&  \frac{2}{\pi \left({\bf q}_1 - {\bf q}_2\right)^2}
\Bigg\{ \frac{1}{2}\ln{\frac{{\bf q}_1^2}{{\bf q}_2^2}} +  \ln{\frac{\left({\bf q}_1-{\bf q}_2\right)^2}{{\bf q}_1^2}}\Bigg\},
\end{eqnarray}
which again agrees with the result of~\cite{Dixon}.

The azimuthal angle averaged result is
\begin{eqnarray}
\tilde{\cal F}_{2}  &=&  \frac{2}{\pi q_1^2 \left(1 - \rho\right)}
\Bigg\{  \ln{\left(1- \rho\right)}  -\frac{\ln{(\rho)} }{4}\Bigg\}.
\end{eqnarray}

\section{Third iteration}

For the third order case we need to calculate
\begin{eqnarray}
{\cal F}_{3}  =  \sum_{n=-\infty}^\infty \frac{e^{ i n \theta}}{\pi q_2^2}
\int \frac{d \gamma}{2 \pi i} \rho^{1-\gamma+\frac{|n|}{2}} 
 \left(\frac{\Gamma
   \left(\gamma \right)\Gamma (\epsilon ) }{\Gamma \left(\gamma +\epsilon\right)}-H_{|n|-\gamma}   - \frac{1}{\epsilon }\right)^3.
\end{eqnarray}
The contributions with non-zero residue are
\begin{eqnarray}
{\cal F}_{3} =  3 \sum_{n=-\infty}^\infty \frac{e^{ i n \theta}}{\pi q_1^2}
\int \frac{d \gamma}{2 \pi i} \rho^{\frac{|n|}{2}-\gamma} 
 \frac{\Gamma
   \left(\gamma \right)\Gamma (\epsilon ) }{\Gamma \left(\gamma +\epsilon\right)}
   \nonumber\\
   &&\hspace{-7cm}\times 
   \Bigg\{\frac{\Gamma
   \left(\gamma \right)\Gamma (\epsilon ) }{\Gamma \left(\gamma +\epsilon\right)}\left(\frac{\Gamma
   \left(\gamma \right)\Gamma (\epsilon ) }{3\, \Gamma \left(\gamma +\epsilon\right)}-H_{|n|-\gamma}   - \frac{1}{\epsilon }\right)
   +\left(H_{|n|-\gamma}  + \frac{1}{\epsilon }\right)^2 \Bigg\}.
\end{eqnarray}
Applying our general formula in eq.~(\ref{General}) we can write it as
\begin{eqnarray}
{\cal F}_{3}  &=&  \sum_{n=-\infty}^\infty  \sum_{S=0}^\infty \frac{e^{ i n \theta} \rho^{S+\frac{|n|}{2}} }{\pi q_1^2}\Bigg\{
 \frac{1}{2 } {\cal R}_3^{\left[3,2,0\right]}  - 3  {\cal R}_3^{\left[2,1,0\right]}
-3   {\cal R}_3^{\left[2,0,1\right]} +3 {\cal R}_3^{\left[1,0,0\right]}  \nonumber\\
&+& \left(\ln{\frac{1}{\rho}} \right) \left({\cal R}_3^{\left[3,1,0\right]} -3 {\cal R}_3^{\left[2,0,0\right]}  \right)+ \left(\ln{\frac{1}{\rho}} \right)^2 \frac{1}{2}{\cal R}_3^{\left[3,0,0\right]} \Bigg\}.   \label{ff33}
\end{eqnarray}
The coefficients can again be obtained from eq.~(\ref{coeffs}) (requiring somewhat more algebraic manipulation), obtaining
\begin{eqnarray}
{\cal R}_3^{\left[3,2,0\right]}  
&=&\frac{6}{\epsilon ^2}-\frac{18 H_S}{\epsilon }+27 H_S^2+27 \psi ^{(1)}(S+1)-\frac{15 \pi
   ^2}{2};
\end{eqnarray}
\begin{eqnarray}
{\cal R}_3^{\left[2,1,0\right]}
&=&\frac{2}{\epsilon ^2}+\frac{2 H_{S+|n|}-4 H_S}{\epsilon }+4 H_S \left(H_S-H_{S+|n|}\right)+4 \psi
   ^{(1)}(S+1)-\pi ^2;\nonumber
\end{eqnarray}
\begin{eqnarray}
{\cal R}_3^{\left[2,0,1\right]} 
&=& -\psi ^{(1)}(S+|n|+1);
\end{eqnarray}
\begin{eqnarray}
{\cal R}_3^{\left[1,0,0\right]}
&=&\frac{1}{\epsilon ^2}+\frac{2 H_{S+|n|}-H_S}{\epsilon }+H_{S+|n|}^2-2 H_S H_{S+|n|}\nonumber\\
&+&\frac{1}{12} \left(6
   H_S^2+6 \psi ^{(1)}(S+1)-\pi ^2\right);
\end{eqnarray}
\begin{eqnarray}
{\cal R}_3^{\left[3,1,0\right]}
 &=& \frac{3}{\epsilon }-9 H_S;
\end{eqnarray}
\begin{eqnarray}
{\cal R}_3^{\left[2,0,0\right]} 
&=& \frac{1}{\epsilon }+H_{S+|n|}-2 H_S;
\end{eqnarray}
and
\begin{eqnarray}
{\cal R}_3^{\left[3,0,0\right]} &=&  1. 
\end{eqnarray}
Once again we see that both the double and single poles in $\epsilon$
cancel in the expression for ${\cal F}_3$. 
Inserting these expressions into eq.~(\ref{ff33}) we obtain
\begin{eqnarray}
{\cal F}_3 =  \sum_{n=-\infty}^\infty \sum_{S=0}^\infty \frac{3 e^{i n \theta } \rho ^{\frac{|n|}{2}+S} }{\pi 
   q_1^2} 
   \Bigg\{\psi (S+1)
   \bigg(2 \left(H_{S+|n|}+\gamma_E \right)+\ln (\rho )\bigg) && \nonumber\\
   &&\hspace{-11cm}+ \psi (S+|n|+1) \bigg(H_{S+|n|}+\ln (\rho )+3 \gamma_E \bigg)
   + \psi^{(1)}(S+|n|+1)\nonumber\\
   &&\hspace{-11cm}+ \psi (S+1)^2+ \psi^{(1)}(S+1)+ \frac{\ln^2(\rho
   )}{6} + 2 \gamma_E  \ln (\rho )- \frac{\pi ^2}{3}+ 4 \gamma_E^2 \Bigg\} \nonumber\\
    &&  \hspace{-11.5cm}=   \sum_{n=-\infty}^\infty \frac{3 e^{i n \theta } \rho ^{\frac{|n|}{2}} }{\pi 
   q_1^2}\sum_{i=1}^{6} S_i,
\end{eqnarray}
where $\gamma_E = - \psi(1)$. We find six distinct sums ($S_{i}$) which we evaluate now. The 
first sum is 
\begin{eqnarray}
S_1 &=& \sum_{S=0}^\infty  \rho ^{S}
   \psi^{(1)}(S+|n|+1)\nonumber\\
   &=& \frac{1}{1-\rho } 
   \left(\psi ^{(1)}(|n|+1) - \sum_{m=1}^\infty \frac{\rho^{m}}{(|n|+m)^2} \right).
   \end{eqnarray}
The second sum is simple if we use eq.~(\ref{rhoHm2}) to obtain
\begin{eqnarray}
S_2 &=& \sum_{S=0}^\infty  \rho ^{S}
    \psi (S+1)^2  \\
   &=& \frac{1}{1-\rho }
   \Bigg\{\ln (1-\rho ) \left[\ln \left(\frac{1-\rho
   }{\rho }\right)+2 \gamma_E \right]+\frac{\pi ^2}{6}+\gamma_E^2 -{\rm Li}_2(1-\rho )\Bigg\}.
    \nonumber
   \end{eqnarray}
The third sum is easy since it corresponds to $S_1$ with $n=0$: 
   \begin{eqnarray}
S_3 &=& \sum_{S=0}^\infty \rho ^{S}
    \psi^{(1)}(S+1)  ~=~ \frac{1}{1-\rho } \left(\frac{\pi^2}{6}- {\rm Li}_2(\rho )\right).
   \end{eqnarray}
The fourth sum is also very simple:   
\begin{eqnarray}
S_4 &=& \sum_{S=0}^\infty \rho ^{S}
   \left( \frac{\ln^2(\rho
   )}{6} + 2 \gamma_E  \ln (\rho )- \frac{\pi ^2}{3}+ 4 \gamma_E^2 \right)
      \nonumber\\
   &=& \frac{1}{1-\rho}\left( \frac{\ln^2(\rho
   )}{6} + 2 \gamma_E  \ln (\rho )- \frac{\pi ^2}{3}+ 4 \gamma_E^2 \right). 
   \end{eqnarray}
The last two sums    
   \begin{eqnarray}
S_5 &=&  \sum_{S=0}^\infty \rho ^{S}
   \Bigg\{\psi (S+1)
   \bigg[2 \left(H_{S+|n|}+\gamma_E \right)+\ln (\rho )\bigg]  \Bigg\} 
   \end{eqnarray}
and 
\begin{eqnarray}
S_6 &=& \sum_{S=0}^\infty  \rho ^{S}
   \Bigg\{ \psi (S+|n|+1) \left[H_{S+|n|}+\ln (\rho )+3 \gamma_E \right]
    \Bigg\} 
   \end{eqnarray}
   are rather complicated and we include their contribution directly in the total sum of all these 
   terms which is
\begin{eqnarray}
 \label{ThirdOrder}
{\cal F}_3 = \sum_{n=-\infty}^\infty\frac{3 e^{i n \theta } \rho ^{\frac{|n|}{2}} }{\pi  q_1^2 (\rho -1)}
\Bigg\{ (\rho -1) \sum_{m=0}^{\infty}  \rho ^m \left(2 H_m  +  H_{m+|n|} \right) H_{m+|n|} \\
&&\hspace{-11cm}+  \sum_{m=1}^\infty \frac{\rho ^m}{m+|n|} \left(\frac{1}{m+|n|}- \ln (\rho )\right)
+ \frac{\pi^2}{6} - \psi^{(1)}(1+|n|) \nonumber\\
&&\hspace{-11cm}-   \ln (\rho ) H_{|n|}- \ln^2 (1-\rho ) - \frac{\ln^2(\rho ) }{6} + \ln (\rho ) \ln (1-\rho )\Bigg\}.
 \nonumber
\end{eqnarray}
This expression corresponds to the Fourier transform in the azimuthal angle of the equivalent order calculation shown in Ref.~\cite{Dixon}. It allows us to calculate the azimuthal angle averaged result which 
corresponds to $n=0$ and reads
\begin{eqnarray}
\tilde{\cal F}_3  =
\frac{6 }{\pi  q_1^2 (1-\rho )}
\Bigg\{ \frac{\ln^2(\rho )}{12}+2 \ln^2(1-\rho )-\ln (\rho ) \ln (1-\rho) + {\rm Li}_2(\rho )\Bigg\}.
\end{eqnarray} 
Note that the dilogarithm appears at this order because the overall factor with the propagator is included when 
performing the azimuthal angle averaging.  

Although lengthy, it is possible to perform the sum over $n$ in eq.~(\ref{ThirdOrder}). 
The first step is to use the relations in eq.~(\ref{Hmn}) and eq.~(\ref{rhoH2mn}) to write the 
sum of the ${\cal S}_i$ in the form
\begin{eqnarray}
\sum_{i=1}^{6} S_i = \frac{3 e^{i n \theta } \rho ^{\frac{|n|}{2}} }{\pi  q_1^2 (\rho -1)}
\Bigg\{ (\rho -1)  \sum_{m=1}^{\infty} \left( \sum_{k=1}^m \frac{2 \rho ^m H_m}{k+|n|} 
+  \sum_{p=0}^\infty  \sum_{s=1}^{p+m} \frac{\rho^{p+m}}{ (m+|n|)(s+|n|)} \right) &&
\nonumber\\
&&\hspace{-14.5cm}+  \sum_{m=1}^\infty \frac{\rho ^m}{m+|n|} \left(\frac{1}{m+|n|}- \ln \rho \right) 
- H_{|n|} \left(H_{|n|} + \ln{ \frac{\rho}{(1-\rho)^2}}  + \sum_{m=1}^\infty \frac{2 \rho^m}{m+|n|}\right)\nonumber\\
&&\hspace{-13cm}
- \psi^{(1)}(1+|n|) + \frac{\pi^2}{6} - \ln^2 (1-\rho ) - \frac{\ln^2\rho  }{6} + \ln \rho  \ln (1-\rho )\Bigg\}. 
\end{eqnarray}
Using this expression we have split the sum over $n$ into six pieces: 
\begin{eqnarray}
{\cal F}_{3}  &=&  \sum_{j=1}^6 {\cal I}_j
\end{eqnarray}
and analytically evaluated each of them. The calculation is tedious and we explain it in Appendix~\ref{sumovernforF3}. We have checked that the sum of our analytic expressions for the ${\cal I}_i$ agrees with our 
numerical implementation of the original integral and the 
corresponding result obtained in Ref.~\cite{Dixon}, {\it i.e.},
\begin{eqnarray}
{\cal F}_3  &=& \frac{6 \ln^2\left(1+\rho-2 \sqrt{\rho } \cos (\theta )\right)
-6 \ln (\rho ) \ln \left(1+\rho-2 \sqrt{\rho } \cos
   (\theta )\right)+\ln^2(\rho )}{2 \pi 
   q_1^2 \left(1+\rho-2 \sqrt{\rho } \cos (\theta )\right)} \nonumber\\
   &&\hspace{-1cm}= \frac{3}{\pi \left({\bf q}_1- {\bf q}_2\right)^2} 
\Bigg\{ \ln^2 {\frac{\left({\bf q}_1- {\bf q}_2\right)^2}{q_1^2}} + \ln{\frac{q_1^2}{q_2^2}} 
\ln {\frac{\left({\bf q}_1- {\bf q}_2\right)^2}{q_1^2}}  + \frac{1}{6} \ln^2{\frac{q_1^2}{q_2^2}}  \Bigg\}.
\end{eqnarray}

\section{A short numerical analysis}

\begin{figure}
\begin{center}
\includegraphics[width=10cm,angle=-90]{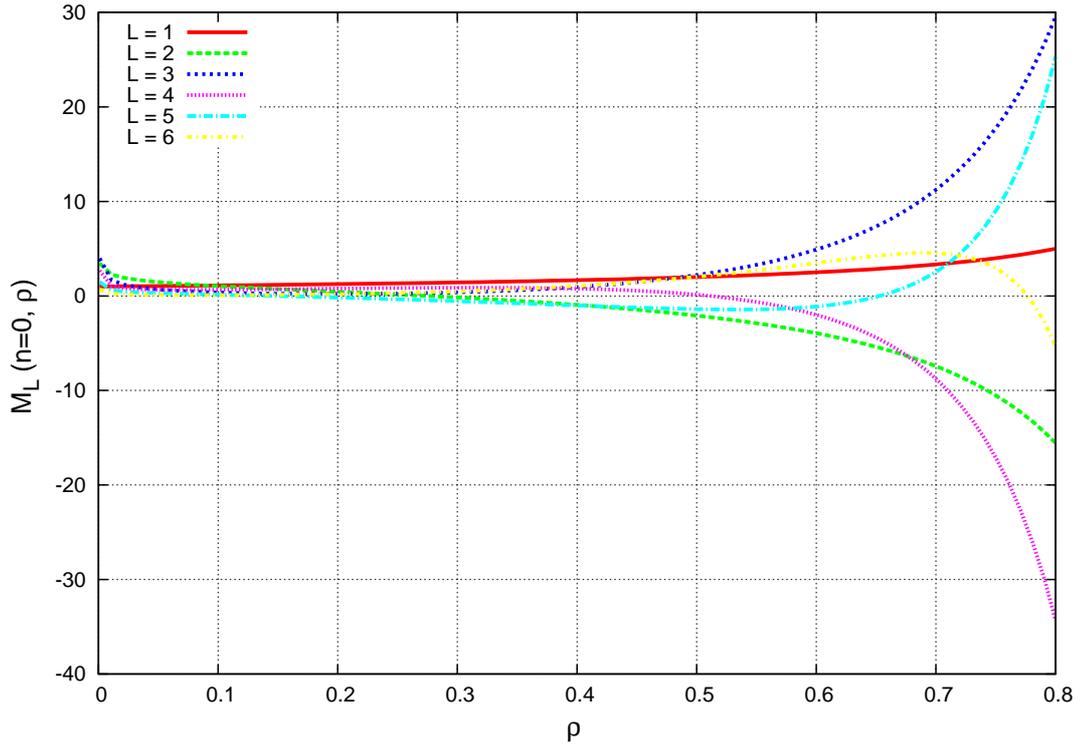}
\end{center}
\caption{Perturbative convergence of the gluon Green function as a function of $\rho$ for $n=0$.}
\label{AverFMn0vsrho}
\end{figure}

We have included the details of the calculation also for ${\cal F}_4$ in Appendix~\ref{F4appendix}. The behavior of 
different Fourier components is not dramatically different from the $n=0$ case which we have studied by defining 
the functions ${\cal M}_L$:
\begin{eqnarray}
\frac{{\cal F}_{L}}{L!}  &=&   \sum_{n=-\infty}^\infty \frac{e^{ i n \theta} \rho^{\frac{|n|}{2}} }{\pi q_1^2}{\cal M}_L (n,\rho).
\end{eqnarray}
We show the first six of them in Fig.~\ref{AverFMn0vsrho}. We observe a good perturbative convergence of these 
functions as $L$ increases for all values of $\rho$.   

\section{Conclusions}

We have presented an analytic procedure to calculate the order-by-order expansion of the gluon Green function in the BFKL 
formalism. Our main result is eq.~(\ref{General}) which presents each term of the expansion as a sum over an infinite number of 
residues of multiple poles which are generated due to integer powers of gamma functions. This structure, which is similar to that found in Mellin-Barnes integrations, arises 
due to our prescription to regularize the gluon propagators with a cut-off, introduced in eq.~(\ref{FMfirstintegral}).

With the method of calculation  described here  we have presented the Fourier transform in the 
azimuthal angle of the results in Ref.~\cite{Dixon}, with which we find agreement up to the orders we have explicitly checked. In detail, the  worked out examples have been 
\begin{eqnarray}
{\cal F}_{1} &=& \sum_{n=-\infty}^\infty 
\frac{ e^{i n \theta } \left(\frac{q_2^2}{q_1^2}\right)^{\frac{|n|}{2}} }{\pi  \left(q_1^2- q_2^2\right)} 
= \frac{1}{\pi \left({\bf q}_1 - {\bf q}_2\right)^2}; 
\end{eqnarray}
\begin{eqnarray}
{\cal F}_{2} &=&  \sum_{n=-\infty}^\infty 2 \frac{e^{i n \theta }
    \left(\frac{q_2^2}{q_1^2}\right)^{\frac{|n|}{2}} }{\pi  \left(q_1^2- q_2^2\right)} \Bigg\{
     \ln \left(1- \frac{q_2^2}{q_1^2} \right)  - \frac{\ln \left(\frac{q_2^2}{q_1^2}\right)}{2}
     - H_{|n|} - \sum_{m=1}^\infty \frac{\left(\frac{q_2^2}{q_1^2}\right)^m}{m+|n|}  
    \Bigg\} \nonumber\\
   &=&  \frac{2}{\pi \left({\bf q}_1 - {\bf q}_2\right)^2}
\Bigg\{ \frac{1}{2}\ln{\frac{{\bf q}_1^2}{{\bf q}_2^2}} +  \ln{\frac{\left({\bf q}_1-{\bf q}_2\right)^2}{{\bf q}_1^2}}\Bigg\};
\end{eqnarray}
\begin{eqnarray}
{\cal F}_{3} &=&  \sum_{n=-\infty}^\infty 
 \frac{3 e^{i n \theta } \left(\frac{q_2^2}{q_1^2}\right)^{\frac{|n|}{2}}  }{\pi  \left(q_1^2- q_2^2\right)}
\Bigg\{ \left(\frac{q_2^2}{q_1^2} -1\right) \sum_{m=0}^{\infty}  \left(\frac{q_2^2}{q_1^2}\right)^m \left(2 H_m  +  H_{m+|n|} \right) H_{m+|n|}\nonumber\\
&+&  \sum_{m=1}^\infty \frac{\left(\frac{q_2^2}{q_1^2}\right)^m}{m+|n|} \left(\frac{1}{m+|n|}- \ln \left(\frac{q_2^2}{q_1^2}\right)\right)
+ \frac{\pi^2}{6} - \psi^{(1)}(1+|n|) \nonumber\\
&-&   \ln \left(\frac{q_2^2}{q_1^2}\right) H_{|n|}- \ln^2 \left(1-\frac{q_2^2}{q_1^2}\right)- \frac{\ln^2\left(\frac{q_2^2}{q_1^2}\right) }{6} + \ln \left(\frac{q_2^2}{q_1^2}\right) \ln \left(1-\frac{q_2^2}{q_1^2}\right)\Bigg\}
 \nonumber\\
   &=& \frac{3}{\pi \left({\bf q}_1- {\bf q}_2\right)^2} 
\Bigg\{ \ln^2 {\frac{\left({\bf q}_1- {\bf q}_2\right)^2}{q_1^2}} + \ln{\frac{q_1^2}{q_2^2}} 
\ln {\frac{\left({\bf q}_1- {\bf q}_2\right)^2}{q_1^2}}  + \frac{1}{6} \ln^2{\frac{q_1^2}{q_2^2}}  \Bigg\}.
\end{eqnarray}
Our approach is general and applicable to any order in the perturbative expansion of the 
gluon Green function for different kernels. 

We will investigate the connection of our general formula~(\ref{General}) with the Knizhnik-Zamolodchikov equations in future 
publications (see a discussion on this in Ref.{\cite{Dixon}). We are also studying the application of this formalism to find the structure of the order-by-order solutions of the 
NLO BFKL equation in different colour representations and its relation to the integrability properties present in scattering amplitudes with 
several Reggeon exchanges.

\subsubsection*{Acknowledgments}
The authors are grateful to James Drummond for useful conversations
and to the Theory Division at CERN, where this work was completed. ASV thanks 
L. {\' A}lvarez-Gaum{\'e}, I. Balitsky, G. Chachamis, A. Kotikov, L. Lipatov and M. A. V{\'a}zquez-Mozo for 
interesting discussions. DAR 
thanks the Leverhulme foundation for financial support.  ASV acknowledges support from the 
European Commission under contract LHCPhenoNet (PITN-GA-2010-264564), the 
Madrid Regional Government (HEPHA- COS ESP-1473), the Spanish Government
(MICINN (FPA2010-17747)) and the Spanish MINECO Centro de Excelencia Severo
Ochoa Programme (SEV-2012-0249).

\appendix

\section{Some useful relations involving sums of harmonic numbers}

\begin{eqnarray}
\sum_{S=0}^\infty \rho^S H_{S} = - \frac{\ln{(1-\rho)}}{1-\rho};
\label{rhoH}
\end{eqnarray}

\begin{eqnarray}
\sum_{S=0}^\infty \rho^S H_{S+|n|} =  \frac{1}{1-\rho}  \left( H_{|n|} 
+  \sum_{S=1}^\infty  \frac{\rho^S}{S+|n|}\right);
\label{rhoHsplusn}
\end{eqnarray}

\begin{eqnarray}
\sum_{m=1}^\infty \rho^m H_{m}^2 = \frac{\zeta(2)+\ln{(1-\rho)} \ln{\left(\frac{1-\rho}{\rho}\right)} - {\rm Li}_2 (1-\rho)}{1-\rho};
\label{rhoHm2}
\end{eqnarray}

\begin{eqnarray}
H_{m+|n|} &=& H_{|n|} + \sum_{k=1}^m \frac{1}{k + |n|};
\label{Hmn}
\end{eqnarray}

\begin{eqnarray}
\sum_{m=1}^\infty \rho^m H_{m+|n|}^2 &=& \frac{\rho}{1-\rho} H_{|n|}^2
+ 2 H_{|n|}  \sum_{m=1}^\infty  \sum_{k=1}^m \frac{\rho^m}{k+|n|} \nonumber\\
&+&\sum_{m=1}^\infty  \sum_{k=1}^m \sum_{s=1}^m \frac{\rho^m}{(k+|n|)(s+|n|)}
\nonumber\\
 &&\hspace{-3cm}= \frac{\rho H_{|n|}^2}{1-\rho} 
+  \frac{2 H_{|n|} }{1-\rho} \sum_{m=1}^\infty \frac{\rho^m}{m+|n|} 
+ \sum_{p=0}^\infty  \sum_{q=1}^\infty \sum_{s=1}^{p+q} \frac{\rho^{p+q}}{(q+n)(s+n)}.
\label{rhoH2mn}
\end{eqnarray}

\section{The sum over $n$ for ${\cal F}_2$}

\begin{eqnarray}
- \sum_{n=-\infty}^\infty 2 \frac{e^{i n \theta }
    \rho ^{\frac{|n|}{2}}}{\pi  q_1^2 (1-\rho)} 
     H_{|n|} &=& 
    \frac{2  }{\pi q_1^2
   \left(1-\rho\right)} \nonumber\\
   &&\hspace{-2cm}\times  \left(\frac{\ln
   \left(1-\sqrt{\rho} 
   e^{i \theta }\right)}{1- \sqrt{\rho} 
   e^{i \theta }} +\frac{ \ln
   \left(1-\sqrt{\rho} e^{- i \theta }\right)}{1
   - \sqrt{\rho} e^{- i \theta }}\right);
   \label{F2n1}
\end{eqnarray}

\begin{eqnarray}
 \sum_{n=-\infty}^\infty 2 \frac{e^{i n \theta } \rho ^{\frac{|n|}{2}}}{\pi  q_1^2 (1-\rho)} \bigg( \ln (1-\rho )  - \frac{\ln (\rho )}{2} \bigg) &&\nonumber\\
 &&\hspace{-3cm}= \frac{2 \ln (1-\rho )-\ln (\rho )}{\pi  q_1^2 \left(1+\rho-2 \sqrt{\rho } \cos
   (\theta )\right)};
     \label{F2n2}
\end{eqnarray}

\begin{eqnarray}
  - \sum_{n=-\infty}^\infty 2 \frac{e^{i n \theta } \rho ^{\frac{|n|}{2}}}{\pi  q_1^2 (1-\rho)} \sum_{m=1}^\infty \frac{\rho^m}{m+|n|} &&\nonumber\\
  &&\hspace{-5cm}= - \frac{2}{\pi  q_1^2 (1-\rho)}  \sum_{m=1}^\infty \left(\frac{\rho^{m}}{m } 
+ 2 \sum_{n=1}^\infty 
     \frac{ \cos{( n \theta) } }{m + n} \rho^{m+\frac{n}{2}}
     \right) \nonumber\\
     &&	\hspace{-5cm} = \frac{2 }{\pi  q_1^2 (1-\rho ) \left(1+\rho-2 \sqrt{\rho } \cos (\theta )\right)}
   \Bigg\{\left(e^{i \theta } \sqrt{\rho }-\rho \right) \ln \left(1-e^{-i \theta } \sqrt{\rho }\right)\nonumber\\
   &&\hspace{-4cm}+\left(e^{-i \theta } \sqrt{\rho }-\rho\right) \ln \left(1-e^{i \theta }
   \sqrt{\rho }\right)+(\rho -1) \ln (1-\rho )\Bigg\}.
     \label{F2n3}
\end{eqnarray}

\section{The sum over $n$ for ${\cal F}_3$}
\label{sumovernforF3}

For ${\cal I}_1$ we have
\begin{eqnarray}
{\cal I}_1 &=&\sum_{n=-\infty}^\infty  \frac{3 e^{i n \theta } \rho ^{\frac{|n|}{2}} }{\pi  q_1^2}
  \sum_{m=1}^{\infty}  \sum_{k=1}^m \frac{2 \rho ^m H_m}{k+|n|}. 
\end{eqnarray}
In this type of sums it is convenient to single out the $n=0$ term:
\begin{eqnarray}
{\cal I}_1 &=& \frac{6}{\pi  q_1^2}
  \sum_{m=1}^{\infty}  \sum_{k=1}^m \frac{\rho ^m H_m}{k} \\
  &+& \sum_{n=1}^\infty  \frac{3 e^{i n \theta } \rho ^{\frac{n}{2}} }{\pi  q_1^2}
  \sum_{m=1}^{\infty}  \sum_{k=1}^m \frac{2 \rho ^m H_m}{k+n}
  +\sum_{n=1}^\infty  \frac{3 e^{-i n \theta } \rho ^{\frac{n}{2}} }{\pi  q_1^2}
  \sum_{m=1}^{\infty}  \sum_{k=1}^m \frac{2 \rho ^m H_m}{k+n}.  \nonumber
\end{eqnarray}
The first line corresponds to 
\begin{eqnarray}
\frac{6}{\pi q_1^2} \sum_{m=1}^\infty \rho^m H_m^2 
\end{eqnarray}
and can be written in the form
\begin{eqnarray}
\frac{6}{\pi q_1^2 (1-\rho)} \Bigg\{\zeta(2)+\ln{(1-\rho)} \ln{\left(\frac{1-\rho}{\rho}\right)} - {\rm Li}_2 (1-\rho)
\Bigg\}.
\end{eqnarray}
The remaining terms can be handled if we introduce
\begin{eqnarray}
\frac{\left(e^{\pm i \theta} \sqrt{\rho}\right)^n}{k +n} &=& \int_0^{e^{\pm i \theta} \sqrt{\rho}} dz 
\frac{z^{k+n-1}}{\left(e^{\pm i \theta} \sqrt{\rho}\right)^k}.
\end{eqnarray}
The final result for this sum can be written in the form
\begin{eqnarray}
{\cal I}_1 &=& \frac{1}{\pi  q_1^2 (\rho -1) \left(1+\rho-2 \sqrt{\rho } \cos (\theta )\right)}
\bigg\{\pi ^2 \left(4 \sqrt{\rho } \cos (\theta )+\rho -1\right) \nonumber\\
&-& 6 \sqrt{\rho } \cos (\theta )
   \ln^2(1-\rho )+6 \ln (1-\rho ) \bigg(\left(1-e^{i \theta } \sqrt{\rho }\right) \ln
   \left(1-e^{-i \theta } \sqrt{\rho }\right)  \nonumber\\
&+&  \left(1-e^{-i \theta } \sqrt{\rho }\right)
   \ln \left(1-e^{i \theta } \sqrt{\rho }\right)+2 i \pi  \sqrt{\rho } \cos (\theta
   )\bigg)
   -6 (\rho -1) \nonumber\\
   &\times& \bigg[ {\rm Li}_2(1-\rho )+{\rm Li}_2\left(\frac{1}{1-e^{i \theta } \sqrt{\rho
   }}\right) +{\rm Li}_2\left(\frac{1}{1-e^{-i \theta } \sqrt{\rho
   }}\right)\nonumber\\
   &-& {\rm Li}_2\left(\frac{\rho -1}{e^{i \theta } \sqrt{\rho
   }-1}\right) -  {\rm Li}_2\left(\frac{\rho -1}{e^{-i \theta } \sqrt{\rho
   }-1}\right)\bigg] \nonumber\\
   &-&12 \sqrt{\rho } \cos (\theta) 
   \left({\rm Li}_2\left(\frac{1}{1-\rho }\right)+ {\rm Li}_2(1-\rho )\right) \bigg\}.
\end{eqnarray}
For the second sum
\begin{eqnarray}
{\cal I}_2 &=& \sum_{n=-\infty}^\infty \frac{3 e^{i n \theta } \rho ^{\frac{|n|}{2}} }{\pi  q_1^2} \sum_{m=1}^{\infty}  
\sum_{p=0}^\infty  \sum_{s=1}^{p+m} \frac{\rho^{p+m}}{ (m+|n|)(s+|n|)}
\end{eqnarray}
we again first work with the $n=0$ terms:
\begin{eqnarray}
\frac{3}{\pi  q_1^2} \sum_{m=1}^{\infty}  
\sum_{p=0}^\infty  \sum_{s=1}^{p+m} \frac{\rho^{p+m}}{ m \, s} &&\nonumber\\
&&\hspace{-3cm}= \frac{\pi ^2-6 {\rm Li}_2\left(\frac{1}{1-\rho
   }\right)-9 \ln^2(1-\rho )+6 \ln (-\rho ) \ln
   (1-\rho )}{2 \pi  q_1^2 (\rho -1)}.
\end{eqnarray}
For the sums with $n \neq 0$ we can make use of the relation 
 \begin{eqnarray}
 \sum_{q=1}^\infty  \frac{\rho^{q}}{q+n} &=& - \ln{(1-\rho)} + n \int_0^1 \frac{d z}{z} z^n \ln{(1-z \rho)} 
 \label{integraltrick}
 \end{eqnarray}
to write
\begin{eqnarray}
\sum_{n=1}^\infty  6 \frac{\cos{( n \theta) }
   \rho^{\frac{n}{2}}
   }{\pi  q_1^2}  \sum_{q=1}^\infty  \frac{\rho^{q}}{q+n} \sum_{p=0}^\infty  \rho^{p} \sum_{s=1}^{p+q} 
   \frac{1}{s+n} && \\
&&\hspace{-8.1cm} =   \sum_{n=1}^\infty  3 \frac{\left(e^{i \theta}
   \sqrt{\rho}\right)^n
   }{\pi  q_1^2}  \sum_{q=1}^\infty   \frac{\rho^{q}}{q+n} 
   \int_0^1 \frac{dx \, x^n}{1-x} 
   \left(\frac{1}{1-\rho}  - \frac{x^{q}}{1-x \rho}\right) + \bigg[\theta \leftrightarrow - \theta\bigg]
   \nonumber\\
&&\hspace{-8.1cm} =      \frac{3 }{\pi  q_1^2}  
   \int_0^1 \frac{dx }{1-x} \int_0^1 \frac{d z}{z} 
   \left(      \frac{\ln{(1-z \rho)}}{1-\rho}  -   
   \frac{\ln{(1-z x \rho)}}{1-x \rho}\right) \sum_{n=1}^\infty   \left(e^{i \theta}
   \sqrt{\rho} x z \right)^n n 
    \nonumber\\
&&\hspace{-7.5cm} - \frac{3 }{\pi  q_1^2}
   \int_0^1 \frac{dx}{1-x} 
   \left(\frac{ \ln{(1-\rho)} }{1-\rho}  - \frac{ \ln{(1-x \rho)} }{1-x \rho}\right)
   \sum_{n=1}^\infty  \left(e^{i \theta}
   \sqrt{\rho} x \right)^n + \bigg[\theta \leftrightarrow - \theta\bigg]
  \nonumber\\
&&\hspace{-8.1cm} =      \frac{3 }{\pi  q_1^2} \int_0^1 \frac{dx }{1-x}  \Bigg\{ 
   \int_0^1 d z 
   \left(      \frac{\ln{(1-z \rho)}}{1-\rho}  -   
   \frac{\ln{(1-z x \rho)}}{1-x \rho}\right) \frac{x  e^{i \theta } \sqrt{\rho }}{\left(1-x z e^{i \theta }
   \sqrt{\rho }\right)^2}
    \nonumber\\
&&\hspace{-7.5cm} +  
   \left(\frac{ \ln{(1-\rho)} }{1-\rho}  - \frac{ \ln{(1-x \rho)} }{1-x \rho}\right)
   \left(1-\frac{1}{1-x e^{i \theta } \sqrt{\rho }} \right) \Bigg\}+ \bigg[\theta \leftrightarrow - \theta\bigg]
   \nonumber\\
&&\hspace{-8.1cm} =      \frac{3 }{\pi  q_1^2} \int_0^1 \frac{dx }{1-x}  \Bigg\{ 
  \frac{\sqrt{\rho } \ln \left(x e^{i \theta } \sqrt{\rho
   }-1\right)}{(1-\rho) \left(\sqrt{\rho }-x e^{i \theta
   }\right)}
   -\frac{\sqrt{\rho } \ln \left(x e^{i \theta }\sqrt{\rho }-1\right)}{(1-x \rho) \left(\sqrt{\rho }-e^{i \theta
   }\right)} +\frac{\ln (1-\rho)}{1-\rho}
   \nonumber\\
   &&\hspace{-7.5cm}
   -\frac{\sqrt{\rho } \ln (\rho -1)}{(1-\rho)
   \left(\sqrt{\rho }-x e^{i \theta }\right)}
   +\frac{\sqrt{\rho } \ln(x \rho -1)}{(1-x \rho) \left(\sqrt{\rho }-e^{i \theta}\right)}
   -\frac{\ln (1-x \rho )}{1-x \rho}
   \Bigg\}+ \bigg[\theta \leftrightarrow - \theta\bigg]  \nonumber
   \end{eqnarray}
   which can be written in the form
   \begin{eqnarray}
  \frac{3 \sqrt{\rho } }{2 \pi  q_1^2 (\rho -1)
   \left(e^{i \theta }-\sqrt{\rho }\right)}
   \Bigg\{
   \left( 
   2 {\rm Li}_2\left(\frac{1}{1-\frac{e^{i \theta}}{\sqrt{\rho }}}\right)
   +2 {\rm Li}_2\left(\frac{e^{i \theta }\sqrt{\rho }-1}{\rho -1}\right)
   \right.
   &&\\
   &&\hspace{-10cm}  -2 {\rm Li}_2\left(\frac{1}{1-\rho }\right) 
   -2 {\rm Li}_2\left(\frac{\sqrt{\rho }(\rho -1)}{e^{i \theta }-\sqrt{\rho }}+\rho \right)
   \nonumber\\
   && \hspace{-10cm}
   +\ln (\rho -1) \left(\ln (\rho )-2 \ln
   \left(\frac{e^{i \theta }-\sqrt{\rho }}{\rho -1}\right)\right)
   \nonumber\\
   && \hspace{-11cm}
   +\ln   \left(e^{i \theta } \sqrt{\rho }-1\right) \left(\ln (\rho )+2
   \ln \left(\frac{1-e^{-i \theta } \sqrt{\rho }}{1-\rho
   }\right)+2 \ln \left(\frac{\sqrt{\rho }-e^{i \theta }}{\rho
   -1}\right)\right)
   \nonumber\\
   && \hspace{-11cm}
   \left.
   +2 \ln (1-\rho ) \ln \left(\frac{1}{\rho
   -1}\right)+2 \pi  (\pi -i \ln (\rho )) -2 i \pi  \ln \left(1-e^{-i \theta } \sqrt{\rho }\right)\right)\nonumber\\
   && \hspace{-11cm}
   -e^{i \theta } \left((\pi +3 i \ln (1-\rho ))^2+6
   \ln (1-\rho ) \ln (\rho )   \right)-6 \, 
   {\rm Li}_2\left(\frac{1}{1-\rho }\right)
   \Bigg\} + {\rm c. c.}\nonumber
 \end{eqnarray}
The third sum reads
\begin{eqnarray}
{\cal I}_3 &=& \sum_{n=-\infty}^\infty  \frac{3 e^{i n \theta } \rho ^{\frac{|n|}{2}} }{\pi  q_1^2 (\rho -1)}
\sum_{m=1}^\infty \frac{\rho ^m}{m+|n|} \left(\frac{1}{m+|n|}- \ln \rho \right) \\
&&\hspace{-1cm}= \frac{3 }{\pi  q_1^2 (\rho -1)}
\sum_{m=1}^\infty \frac{\rho ^m}{m} \left(\frac{1}{m}- \ln \rho \right) \nonumber\\
&&\hspace{-.8cm}+ \sum_{n=1}^\infty  \frac{3 \left(e^{i n \theta }+e^{-i n \theta }\right) \rho ^{\frac{n}{2}} }{\pi  q_1^2 (\rho -1)}
\sum_{m=1}^\infty \frac{\rho ^m}{m+n} \left(\frac{1}{m+n}- \ln \rho \right) \nonumber\\
&&\hspace{-1cm}=  \frac{3 }{\pi  q_1^2 (\rho -1)}
\bigg\{{\rm Li}_2(\rho )+\ln (1-\rho ) \ln (\rho) \nonumber\\
&&\hspace{-.8cm}+\frac{{\rm Li}_2(\rho
   )+\ln (1-\rho ) \ln (\rho )-e^{i \theta } \sqrt{\rho }
   \left({\rm Li}_2\left(e^{-i \theta } \sqrt{\rho
   }\right)+\ln (\rho ) \ln \left(1-e^{-i \theta
   } \sqrt{\rho }\right)\right)}{e^{i \theta }
   \sqrt{\rho }-1} \nonumber\\
&&\hspace{-.8cm}+\frac{
   {\rm Li}_2(\rho )+\ln (1-\rho ) \ln (\rho)-e^{-i \theta }\sqrt{\rho }
   \left({\rm Li}_2\left(e^{i \theta } \sqrt{\rho
   }\right)+\ln (\rho ) \ln \left(1-e^{i \theta }
   \sqrt{\rho }\right)\right)}{e^{-i \theta }\sqrt{\rho }-1}\bigg\}.\nonumber
\end{eqnarray}
For the sum
\begin{eqnarray}
{\cal I}_4 = - \sum_{n=-\infty}^\infty  \frac{3 e^{i n \theta } \rho ^{\frac{|n|}{2}} }{\pi  q_1^2 (\rho -1)}
H_{|n|} \left(H_{|n|} + \ln{ \frac{\rho}{(1-\rho)^2}}  + \sum_{m=1}^\infty \frac{2 \rho^m}{m+|n|}\right)
\end{eqnarray}
we use the result
\begin{eqnarray}
\sum_{n=-\infty}^\infty  \frac{3 e^{i n \theta } \rho^{\frac{|n|}{2}}}{\pi  q_1^2
   \left(1-\rho\right)}  H_{|n|}^2 &&\\
   &&\hspace{-4cm}= \frac{1}{2 \pi  q_1^2 (1-\rho )}
   \Bigg\{\frac{\pi ^2-6
   {\rm Li}_2\left(1-e^{i \theta } \sqrt{\rho
   }\right)+6 \ln \left(\frac{e^{-i \theta
   }}{\sqrt{\rho }}-1\right) \ln \left(1-e^{i \theta
   } \sqrt{\rho }\right)}{1-e^{i \theta }
   \sqrt{\rho }} \nonumber\\
   &&\hspace{-3cm} +\frac{\pi^2-6 {\rm Li}_2\left(1-e^{-i \theta }
   \sqrt{\rho }\right)+6 \ln \left(\frac{e^{i
   \theta }}{\sqrt{\rho }}-1\right) \ln
   \left(1-e^{-i \theta } \sqrt{\rho }\right)}{1-e^{-i \theta } \sqrt{\rho }}\Bigg\} \nonumber
   \end{eqnarray}
and  
\begin{eqnarray}
- \sum_{n=-\infty}^\infty  \frac{3 e^{i n \theta } \rho ^{\frac{|n|}{2}} }{\pi  q_1^2 (\rho -1)}
H_{|n|}  \ln{ \frac{\rho}{(1-\rho)^2}} &&\\
&&\hspace{-5.5cm}= \frac{3 }{\pi  q_1^2 (\rho -1)}
\ln \left(\frac{\rho }{1-\rho ^2}\right)
   \Bigg\{\frac{\ln \left(1-e^{i \theta }
   \sqrt{\rho }\right)}{e^{i \theta } \sqrt{\rho}-1}+\frac{\ln \left(1-e^{-i \theta }
   \sqrt{\rho }\right)}{e^{-i \theta }
   \sqrt{\rho }-1}\Bigg\}. \nonumber
\end{eqnarray}  
We also need the integral representation in  eq.~(\ref{integraltrick}) to write
\begin{eqnarray}
 \sum_{n=-\infty}^\infty \frac{6 e^{i n \theta } \rho ^{\frac{|n|}{2}}}{ \pi  q_1^2 (1-\rho)} 
H_{|n|}  \sum_{m=1}^\infty \frac{\rho^m}{m + |n|} &&\\
&&\hspace{-6cm}= \frac{6 }{ \pi  q_1^2 (1-\rho)} \Bigg\{
 \int_0^1 	\frac{d z}{z}  \ln{(1-z \rho)}
\sum_{n=1}^\infty n \left(e^{i \theta } \sqrt{\rho} z \right)^n H_{n}
 \nonumber\\
&&\hspace{-3cm}- \ln{(1-\rho)}
 \sum_{n=1}^\infty  \left(e^{i \theta } \sqrt{\rho}\right)^n
H_{n}  \Bigg\}   + 
 \bigg[\theta \leftrightarrow - \theta\Bigg]
 \nonumber\\
&&\hspace{-6cm}= \frac{6 }{ \pi  q_1^2 (1-\rho)} \Bigg\{ 
e^{i \theta } \sqrt{\rho} \int_0^1 	d z \ln{(1-z \rho)}
  \frac{\left(1-\ln{\left(1-e^{i \theta } \sqrt{\rho} z\right)}\right)}{\left(1-e^{i \theta } \sqrt{\rho} z\right)^2} 
\nonumber\\
&&\hspace{-3cm}+ \ln{(1-\rho)}
  \frac{\ln{\left(1-e^{i \theta } \sqrt{\rho}\right)}}{1-e^{i \theta } \sqrt{\rho}}
  \Bigg\}   + 
 \bigg[\theta \leftrightarrow - \theta\bigg] \nonumber\\
&&\hspace{-6cm}=
  \frac{6 \sqrt{\rho }}{\pi 
   q_1^2 (1-\rho ) \left(e^{i
   \theta }-\sqrt{\rho }\right)}  \Bigg\{ 
   {\rm Li}_2\left(\frac{1}{1-\frac{e^{i \theta
   }}{\sqrt{\rho }}}\right)-
   {\rm Li}_2\left(\frac{\sqrt{\rho } (\rho -1)}{e^{i
   \theta }-\sqrt{\rho }} +\rho \right)\nonumber\\
   &&\hspace{-5cm}+\ln
   \left(1-e^{i \theta } \sqrt{\rho }\right) \left(\frac{1}{2}\ln
   \left(1-e^{i \theta } \sqrt{\rho }\right)- \ln
   \left(\frac{e^{i \theta } (\rho -1)}{\sqrt{\rho
   }-e^{i \theta }}\right)\right)\Bigg\} \nonumber\\
    &&\hspace{-5cm} +  \bigg[\theta \leftrightarrow - \theta\bigg].  \nonumber
    \end{eqnarray}         
The fifth sum is rather simple:   
\begin{eqnarray}
{\cal I}_5 &=& -  \sum_{n=-\infty}^\infty  \frac{3 e^{i n \theta } \rho ^{\frac{|n|}{2}} }{\pi  q_1^2 (\rho -1)}
\psi^{(1)}(1+|n|) \\
&&\hspace{-1.4cm}= \frac{1}{2 \pi  q_1^2 (1-\rho )} \Bigg\{\frac{\pi ^2
   e^{i \theta } \sqrt{\rho }-6 \, {\rm Li}_2\left(e^{i
   \theta } \sqrt{\rho }\right)}{1-e^{i \theta }
   \sqrt{\rho }}+\frac{\pi ^2 e^{-i \theta } \sqrt{\rho }-6 \, 
  {\rm Li}_2\left(e^{-i \theta } \sqrt{\rho
   }\right)}{1-e^{-i \theta } \sqrt{\rho }}+\pi ^2\Bigg\}.\nonumber
\end{eqnarray}
The sixth, and last sum, is also very easy:
\begin{eqnarray}
{\cal I}_6 &=& \sum_{n=-\infty}^\infty  \frac{3 e^{i n \theta } \rho ^{\frac{|n|}{2}} }{\pi  q_1^2 (\rho -1)}
\left(\frac{\pi^2}{6} - \ln^2 (1-\rho ) - \frac{\ln^2\rho  }{6} + \ln \rho  \ln (1-\rho )\right) \nonumber\\
&=& \frac{-3 \left(\frac{\pi^2}{6} - \ln^2 (1-\rho ) - \frac{\ln^2\rho  }{6} + \ln \rho  \ln (1-\rho )\right)  }{\pi  q_1^2 \left(1-e^{i \theta}\sqrt{\rho }\right) \left(1-e^{-i \theta } \sqrt{\rho}\right)}. 
\end{eqnarray}

\section{Representation of ${\cal F}_4$}
\label{F4appendix}

\begin{eqnarray}
{\cal F}_{4}  &=&  \sum_{n=-\infty}^\infty  \sum_{S=0}^\infty \frac{e^{ i n \theta} \rho^{S+\frac{|n|}{2}} }{\pi q_1^2}\Bigg\{
\frac{1}{6} \ln ^3\left(\frac{1}{\rho }\right) 
{\cal R}_4^{\left[4,0,0\right]}\nonumber\\
&+&\ln^2\left(\frac{1}{\rho }\right) \Bigg[-2 
{\cal R}_4^{\left[3,0,0\right]} +\frac{1}{2} {\cal R}_4^{\left[4,0,1\right]}
+\frac{1}{2} 
{\cal R}_4^{\left[4,1,0\right]}\Bigg]\nonumber\\
&+&\ln \left(\frac{1}{\rho }\right)\Bigg[6  
{\cal R}_4^{\left[2,0,0\right]}-4 
{\cal R}_4^{\left[3,0,1\right]}-4 
{\cal R}_4^{\left[3,1,0\right]}+\frac{1}{2}
 {\cal R}_4^{\left[4,0,2\right]}\nonumber\\
&+&
{\cal R}_4^{\left[4,1,1\right]}+\frac{1}{2} 
{\cal R}_4^{\left[4,2,0\right]}\Bigg]-4{\cal R}_4^{\left[1,0,0\right]}+6
{\cal R}_4^{\left[2,0,1\right]}\nonumber\\
&+&6
{\cal R}_4^{\left[2,1,0\right]}-2{\cal R}_4^{\left[3,0,2\right]}-4
  {\cal R}_4^{\left[3,1,1\right]}-2{\cal R}_4^{\left[3,2,0\right]}+\frac{1}{6}
  {\cal R}_4^{\left[4,0,3\right]}+\frac{1}{2}{\cal R}_4^{\left[4,1,2 \right]}\nonumber\\
&+&\frac{1}{2}
  {\cal R}_4^{\left[4,2,1\right]}+\frac{1}{6}{\cal R}_4^{\left[4,3,0\right]} 
 \Bigg\}.
\end{eqnarray}

\begin{eqnarray}
{\cal R}_4^{\left[1,0,0\right]} &=& \frac{1}{12} \left(18 H_S^2 H_{S+|n|}+18 H_{S+|n|} \left(\psi ^{(1)}(S+1)-2 \psi (S+1) H_{S+|n|}\right) \right.\nonumber\\
&+& 12 \psi (S+|n|+1)^3-3 \left(12 \gamma_E ^2+\pi ^2\right) \psi(S+|n|+1)-2 H_S^3 \nonumber\\
   &+& \left. H_S \left(\pi ^2-6 \psi ^{(1)}(S+1)\right)-2 \psi
   ^{(2)}(S+1)-4 \zeta (3)-3 \gamma_E  \left(8 \gamma_E ^2+\pi ^2\right)\right)\nonumber\\
   &+&\frac{1}{\epsilon ^3}+\frac{3
   H_{S+|n|}-H_S}{\epsilon ^2}+\frac{1}{12 \epsilon} \bigg(36 \left(H_{S+|n|}\right){}^2+12 H_S \left(\gamma_E -3
   H_{S+|n|}\right)\nonumber\\
   &+&6 \psi (S+1)^2+6 \psi ^{(1)}(S+1)-\pi ^2-6 \gamma_E ^2\bigg).
\end{eqnarray}
\begin{eqnarray}
{\cal R}_4^{\left[2,0,0\right]} &=& \frac{1}{\epsilon ^2}+\frac{2 H_{S+|n|}-2 H_S}{\epsilon }-4 H_S H_{S+|n|}+
H_{S+|n|}^2+2
 H_S^2\nonumber\\
 &+&\psi ^{(1)}(S+1)-\frac{\pi ^2}{6}.
\end{eqnarray}
\begin{eqnarray}
{\cal R}_4^{\left[2,0,1\right]} = 2 \left(2 H_S-H_{S+|n|}\right) \psi ^{(1)}(S+|n|+1)-\frac{2 \psi ^{(1)}(S+|n|+1)}{\epsilon }.
\end{eqnarray}
\begin{eqnarray}
{\cal R}_4^{\left[2,1,0\right]} &=& \frac{2}{\epsilon ^3}+\frac{4 H_{S+|n|}-4 H_S}{\epsilon
   ^2} \nonumber\\
   &+&\frac{-8 H_S H_{S+|n|}+2 H_{S+|n|}^2+4 H_S^2+4 \psi ^{(1)}(S+1)-\pi
   ^2}{\epsilon }\nonumber\\
   &+& \frac{1}{3} \left(6 \psi (S+1) \left(2 \left(3 \gamma_E -H_{S+|n|}\right) \psi (S+|n|+1) \right.\right.\nonumber\\
   &-& \left. 4
   \psi ^{(1)}(S+1)+\pi ^2+2 \gamma_E ^2\right)+24 \psi (S+1)^2 \psi (S+|n|+1) \nonumber\\
   &-&  6 \psi (S+|n|+1) \left(2 \gamma_E  \psi (S+|n|+1)-4 \psi ^{(1)}(S+1)+\pi ^2\right)\nonumber\\
   &-& \left.8 \psi
   (S+1)^3-5 \psi ^{(2)}(S+1)-4 \zeta (3)+4 \gamma_E ^3\right).
\end{eqnarray}
\begin{eqnarray}
{\cal R}_4^{\left[3,0,0\right]} &=& H_{S+|n|}-3 H_S+\frac{1}{\epsilon }.
\end{eqnarray}
\begin{eqnarray}
{\cal R}_4^{\left[3,0,1\right]} &=& -\psi ^{(1)}(S+|n|+1).
\end{eqnarray}
\begin{eqnarray}
{\cal R}_4^{\left[3,0,2\right]} &=& \psi ^{(2)}(|S|+n+1).
\end{eqnarray}
\begin{eqnarray}
{\cal R}_4^{\left[3,1,0\right]} &=&\frac{3}{\epsilon ^2}+ \frac{3 H_{S+|n|}-9 H_S}{\epsilon } \nonumber\\
&+&\frac{1}{4} \left(-36 H_S H_{S+|n|}+54 H_S^2+30 \psi
   ^{(1)}(S+1)-7 \pi ^2\right).
\end{eqnarray}
\begin{eqnarray}
{\cal R}_4^{\left[3,1,1\right]} &=& 9 H_S \psi ^{(1)}(S+|n|+1)-\frac{3 \psi ^{(1)}(S+|n|+1)}{\epsilon }.
\end{eqnarray}
\begin{eqnarray}
{\cal R}_4^{\left[3,2,0\right]} &=& \frac{6}{\epsilon ^3}+\frac{6 \left(H_{S+|n|}-3 H_S\right)}{\epsilon
   ^2}\nonumber\\
   &+&\frac{3 \left(6 H_S \left(3 H_S-2 H_{S+|n|}\right)+18 \psi ^{(1)}(S+1)-5 \pi ^2\right)}{2
   \epsilon } \nonumber\\
   &-&\frac{3}{2} \left(\left(-36 \gamma_E  H_S+5 \pi ^2+18 \gamma_E ^2\right) H_{S+|n|} \right.\nonumber\\
   &-&  18 \left(\psi
   (S+1)^2+\psi ^{(1)}(S+1)\right) H_{S+|n|}+18 H_S^3\nonumber\\
   &-&\left. 15 \pi ^2 H_S+54 H_S \psi
   ^{(1)}(S+1)+10 \psi ^{(2)}(S+1)+4 \zeta (3)\right).
\end{eqnarray}
\begin{eqnarray}
{\cal R}_4^{\left[4,0,0\right]} &=& 1.
\end{eqnarray}
\begin{eqnarray}
{\cal R}_4^{\left[4,0,1\right]} &=& 0.
\end{eqnarray}
\begin{eqnarray}
{\cal R}_4^{\left[4,0,2\right]}&=& 0.
\end{eqnarray}
\begin{eqnarray}
{\cal R}_4^{\left[4,0,3\right]}&=& 0.
\end{eqnarray}
\begin{eqnarray}
{\cal R}_4^{\left[4,1,0\right]} &=& 4 \left(\frac{1}{\epsilon }-4 H_S\right).
\end{eqnarray}
\begin{eqnarray}
{\cal R}_4^{\left[4,1,1\right]}&=& 0.
\end{eqnarray}
\begin{eqnarray}
{\cal R}_4^{\left[4,1,2\right]}&=& 0.
\end{eqnarray}
\begin{eqnarray}
{\cal R}_4^{\left[4,2,0\right]} &=& \frac{12}{\epsilon ^2}-\frac{48 H_S}{\epsilon }+96 \left(H_S\right){}^2+56 \psi ^{(1)}(S+1)-\frac{44
   \pi ^2}{3}.
\end{eqnarray}
\begin{eqnarray}
{\cal R}_4^{\left[4,2,1\right]} &=& 0.
\end{eqnarray}
\begin{eqnarray}
{\cal R}_4^{\left[4,3,0\right]} &=& \frac{24}{\epsilon ^3}-\frac{96 H_S}{\epsilon ^2}+\frac{8 \left(24 H_S^2+24 \psi
   ^{(1)}(S+1)-7 \pi ^2\right)}{\epsilon }\nonumber\\
   &-&8 \left(4 H_S \left(8 H_S^2+24 \psi ^{(1)}(S+1)-7 \pi ^2\right)+17 \psi ^{(2)}(S+1)\right.\nonumber\\
   &+&\left. 4
   \zeta (3)\right).
\end{eqnarray}

\begin{eqnarray}
{\cal F}_{4}  &=&  \sum_{n=-\infty}^\infty  \sum_{S=0}^\infty \frac{e^{ i n \theta} \rho^{S+\frac{|n|}{2}} }{\pi q_1^2}\Bigg\{\frac{1}{6} \left(12 \left(-3 \psi^2(S+1) \left(2 H_{S+|n|}+\ln (\rho )\right) 
\right. \right.\nonumber\\
&-& 6 \psi
   (S+1) \psi (S+|n|+1) \left(H_{S+|n|}+\ln (\rho )+3 \gamma_E \right)\nonumber\\
&-&\psi (S+|n|+1)
   \left(2 H_{S+|n|}+\ln (\rho )+10 \gamma_E \right) (\psi (S+|n|+1)+\ln (\rho ))\nonumber\\
&-&2 \psi ^{(1)}(S+|n|+1) \left(3
   \left(H_{S+|n|}+H_S\right)+\ln (\rho )\right)\nonumber\\
&-&\left. 2 \psi
   ^{(1)}(S+1) \left(3 H_{S+|n|}+\ln (\rho )\right)-6 H_S \psi ^{(1)}(S+1)-2 \psi (S+1)^3\right)\nonumber\\
&+& 24
   \left(\pi ^2-12 \gamma_E^2\right) \psi (S+|n|+1)-12 \psi ^{(2)}(S+|n|+1)\nonumber\\
&-&12 H_S \left(\ln
   ^2(\rho )+12 \gamma_E  \ln (\rho )-2 \pi ^2+12 \gamma_E^2\right)-72 \gamma  H_S^2-12
   \psi ^{(2)}(S+1)\nonumber\\
&-&\left.\ln (\rho ) \left(\ln^2(\rho )+12 \gamma_E  \ln (\rho )-8 \pi ^2\right)+24 \gamma_E 
   \left(\gamma_E^2+\pi ^2\right)\right) \Bigg\}. 
\end{eqnarray}
The azimuthal angle averaged result is 
\begin{eqnarray}
\tilde{\cal F}_{4}  &=&   \sum_{S=0}^\infty \frac{ \rho^{S} }{\pi q_1^2}\Bigg\{ -\frac{1}{6}\ln^3(\rho )
- 4 H_S \ln^2(\rho )+ \left(\frac{4}{3} \pi ^2 - 24 H_S^2 \right)\ln (\rho ) \\
&-&8 \psi ^{(1)}(S+1)
   \left(6 H_S+\ln (\rho )\right)-32 H_S^3+ 8 \pi ^2 H_S- 4 \psi ^{(2)}(S+1)
\Bigg\}. \nonumber
\end{eqnarray}

\end{document}